\documentclass[aps,pra,floatfix,groupedaddress,nofootinbib]{revtex4}

\usepackage[latin1]{inputenc}
\usepackage[english]{babel}
\usepackage{setspace, graphicx, amssymb}
\usepackage{amsmath}
\usepackage{amscd}
\usepackage{color}
\usepackage{epstopdf}
\usepackage{subfigure}
\usepackage{natbib}
\usepackage{bbold}

\newcommand{\tr}{{\rm Tr\thinspace}}

\newcommand{\ket}[1]{\left\vert{#1}\right\rangle}
\newcommand{\ketil}[1]{\vert{#1}\rangle}

\newcommand{\order}[1]{\mathcal{O}\!\left( #1 \right)}

\newcommand{\N}{\mathcal{N}}

\newcommand{\erf}[1]{Eq.~(\ref{#1})}

\newcommand{\myvcenter}[1]{\ensuremath{\vcenter{\hbox{#1}}}}

\newcommand{\dg}{^\dagger}
\newcommand{\dt}[1]{\textrm{d}{#1}}
\newcommand{\nn}{\nonumber}

\newcommand{\BQIC}{Berkeley Center for Quantum Information and Computation, Berkeley, California 94720 USA}
\newcommand{\DeptPhys}{Department of Physics, University of California, Berkeley, California 94720 USA}
\newcommand{\DeptChem}{Department of Chemistry, University of California, Berkeley, California 94720 USA}
\newcommand{\sandia}{Sandia National Laboratories, Livermore, California 94550 USA}

\begin{document}
	
\title{Finite temperature quantum simulation of stabilizer Hamiltonians}

\author{Kevin C. Young$^{1,2,4}$\footnote{Electronic address: \texttt{kyoung@sandia.gov}}, 
Mohan Sarovar$^{1,3,4}$,
Jon Aytac$^{1,2}$, C. M. Herdman$^{1,2,3}$ and K. Birgitta Whaley$^{1,3}$ \vspace{.3cm}}

\address{$^1$\BQIC}
\address{$^2$\DeptPhys}
\address{$^3$\DeptChem}
\address{$^4$\sandia\vspace{.2cm}}

\date{\today}
\begin{abstract}
	We present a scheme for robust finite temperature quantum simulation of stabilizer Hamiltonians.  The scheme is designed for realization in a physical system consisting of a finite set of neutral atoms trapped in an addressable optical lattice that are controllable via 1- and 2-body operations together with dissipative 1-body operations such as optical pumping. We show that these minimal physical constraints suffice for design of a quantum simulation scheme for any stabilizer Hamiltonian at either finite or zero temperature. We demonstrate the approach with application to the abelian and non-abelian toric codes.  
\end{abstract}
\maketitle 

\section{Introduction} 
\label{sec:introduction}

There has been much recent work on approaches to experimentally engineer many-body quantum phases of matter~\cite{Ioffe2002,Buchler2005,Greiner2002a,Jaksch2005,Micheli2006a,Pupillo2008a,Weimer2010a}. In particular there is a wide array of lattice Hamiltonians whose ground states are novel quantum phases that are not yet known to exist in natural systems~\cite{Wen2003,Moessner2001d,Levin2005a,Kitaev2003,Kitaev2006a}; to physically realize such phases, one must generate the models artificially. Additionally, many of these novel phases have potential applications to quantum information~\cite{Dennis2002,Freedman2002a,Kitaev2003,Nayak2008}. A robust experimental realization of such phases of matter might be a route to building a fault-tolerant quantum computer~\cite{Preskill1998}.

One such class of lattice Hamiltonians are quantum stabilizer Hamiltonians~\cite{Bacon2006,Bombin2011,Bombin2008,Bombin2007,Bombin2006,Bravyi1998,Dennis2002,Kitaev2003,Yoshida2011a,Raussendorf2001}. Stabilizer Hamiltonians are composed of commuting multi-qubit Pauli operators and play a key role in quantum error correction.  Encoding quantum information into ground states of stabilizer Hamiltonians provides a natural physical architecture for realization of quantum error correcting codes and quantum memory in qubit arrays~\cite{Dennis2002}. Because of their characteristic degenerate ground states and finite energy gaps to local excitations, they offer physical protection for encoded quantum information. Much recent interest  has therefore focused on the quantum phases generated by stabilizer Hamiltonians, their ability to generate topological order and their relation to novel exotic phases~\cite{Aguado2008a,Weimer2010a,Verstraete2009a,Brennen2009b}.   

While the quantum codes derived from ground states of stabilizer Hamiltonians are of broad theoretical interest in quantum information theory, the transition from theoretical characterization to actual realization of these in physical systems is nevertheless impeded by a number of difficulties. Many stabilizer Hamiltonians are composed of non-local or many-body terms that are difficult to realize in practice. This usually requires significant effort and associated overhead in quantum engineering of interactions.  Both trapping of ions in Paul traps and of atoms or molecules in optical lattices offer the ability to generate the required many-body interactions. Additionally, a robust simulation of a lattice Hamiltonian requires an entropy sink to remove entropy that accumulates from noise in the control operations and interactions with the environment~\cite{Brown2007,Lloyda,Herdman2010c}. Entropy can be removed either actively via algorithmic error correction, or passively via an effective coupling to an external reservoir~\cite{Schirmer2010,Verstraete2009a,Lloyda,Kraus2008,Diehl2008}. In this work we will show how to overcome these challenges for the robust quantum simulation of both abelian and non-abelian toric code Hamiltonians and generate an effective coupling to a low temperature thermal reservoir.

In the rest of this paper we first provide an introduction to stabilizer Hamiltonians in Section~\ref{sec:stabilizer}.  In Section~\ref{sec:toric_code} we then summarize the properties of the well known abelian and non-abelian toric code Hamiltonians of Kitaev, which constitute key examples of stabilizer Hamiltonians with topologically ordered ground states and abelian and non-abelian excitations, respectively.  We then present a detailed discussion of thermalization for stabilizer Hamiltonians in Section~\ref{sec:thermalization_of_toric_code}.  The analysis given here provides both a generalization and a more efficient approach to generating thermalization than that employed in our previous work~\cite{Herdman2010c}.  This improved approach to thermalization constitutes the main new set of results in this work. Following this, we summarize several routes to simulation of the stabilizer Hamiltonian and components needed for thermalization, including a non-perturbative approach that is considerably more efficient than our previous perturbative stroboscopic approach (Section~\ref{sec:simulating_many_body_quantum_operators}).  We outline how such a thermal quantum simulation of the abelian toric code may be implemented with a finite set of neutral atoms trapped 
in an optical lattice and indicate what is required in order to generalize this to thermal quantum simulation of the non-abelian toric code.  Section~\ref{sec:summary} concludes with a brief summary.

\section{Stabilizer Hamiltonians} 
\label{sec:stabilizer}

Consider a class of lattice systems whose degrees of freedom are $d-$level systems (qudits). We consider Hamiltonians with local $n$-body interactions
\begin{align}
\label{eq:general_stabilizer}
H = -\sum_\nu J_\nu \sum_\N h_\N^\nu
\end{align}
where $\N$ is a spatial neighborhood, $\nu$ labels the type of interaction, and $\{J_\nu\}$ are the interaction strengths with $J_\nu > 0$. We assume that $\{h_\N^\nu\}$ are local $n$-body projection operators with maximum eigenvalue $1$. We will now consider the eigenstates of a local term $h \in \{ h_\N^\nu \}$. We will label the eigenstates of $h$ as $\ket{ \epsilon_d }$, where $ \epsilon > 0$, the eigenvalue of $h$ is $(1-\epsilon)$, and the label $d$ distinguishes degeneracies:
\begin{align}
h \ket{ \epsilon_d } = (1-\epsilon) \ket{ \epsilon_d }.
\end{align}
With this notation, the ground state of each local term in the Hamiltonian is $\ket{0}$, with eigenvalue $1$ and $\epsilon = 0$. We can consider the eigenoperators $\{ b_d,b_d^\dagger \}$ that span the eigenspace of $h_\N^\nu$:
\begin{align}
b_d^\dagger \ket{ \epsilon_c } = \delta_{\epsilon,0} \ket{ \epsilon_d }, &\quad b_d \ket{ \epsilon_c} = \delta_{d,c} \ket{0} \\
&h = \mathbb{1} - \sum_d b_d^\dagger b_d
\end{align}
An arbitrary eigenoperator can be formed from products of $\{ b_d,b_d^\dagger \}$.

We will now consider the case of such Hamiltonians for which all the $h_\N^\nu$ commute.  These constitute the class of stabilizer Hamiltonians, whose ground states include the well known stabilizer codes ~\cite{Bacon2006,Yoshida2011a,Dennis2002}. In this case, eigenstates of $H$ will be simultaneous eigenstates of all $h_\N^\nu$. In particular the ground states of $H$ will be the ground state of all local terms:
\begin{align}
h_\N^\nu \ket{ \Psi_0 } = \ket{\Psi_0}\quad \forall \left(\N,\nu \right)
\end{align}
Excited eigenstates of $H$ can be generate by applying products of $b_d^\dagger$ to the ground state. For example, the state
\begin{align}
\ket{ \Psi_{ex} [ \left( \N, \mu, d \right), \left( \N', \mu, d \right) ] } = b^\dagger_{\N,\mu,d} b^\dagger_{\N',\mu,d} \ket{\Psi_0}
\end{align}
has purely \emph{localized} quasiparticle excitations at neighborhoods $\N$ and $\N'$. Therefore, the eigenoperators of $H$ are consequently purely local, and can be formed from local superpositions of products of the $\{ b^\dagger_{\N,\mu,d} \}$. This locality of the eigenoperators is a key feature of stabilizer-like Hamiltonians.  We expect that an arbitrary translationally invariant local Hamiltonian with non-communting terms will have momentum eigenstates, and therefore the eigenoperators will be extensive superpositions of local operators.

The energy cost of a local quasiparticle excitation  $\left( \N, \mu, a \right)$ is $\Delta E_a^\mu = \epsilon^\mu_a J_\mu$. An arbitrary one-qudit operator acting on qudit $j$, $\sigma_j$, will be formed from a finite local sum of  products of the eigenoperators of all neighborhoods $\N_i, i \in \N_i$. Therefore an arbitrary  one-qudit operator will only create, annihilate or translate quasiparticles within the local region of neighborhoods $\N_i$. In contrast, an arbitrary Hamiltonian with non-commuting local terms will have eigenstates with propagating quasiparticles; consequently a local operator acting on an eigenstates will generically create an non-local superposition of quasiparticles.

\section{Topological Phases and Toric Codes}
\label{sec:toric_code}
Topologically ordered phases of matter are 2D quantum liquid states with no broken conventional symmetry~\cite{Wen1990,Nayak2008}. These topological phases have a quantum ordering which cannot be detected by a local order parameter. The topological order results in a robust ground state degeneracy on surfaces with a non-trivial topology and a finite energy gap to anyonic quasiparticle excitations. Different ground states can only be distinguished by non-local operators that wind around a non-contractible loop of the surface.

Such topological phases have been proposed as the basis for a physically fault tolerant quantum computer~\cite{Freedman2002a,Kitaev2003,Nayak2008,Dennis2002}. Quantum information can be encoded in the degenerate ground states; since all local operators cause transitions to excited state above the finite gap, these phases are relatively insensitive to local perturbations.  Furthermore, both the tunneling amplitude between ground states and splitting of the ground state degeneracy are suppressed exponentially in the system size.  Logical operations can be performed by creating and braiding the anyonic excitations~\cite{Bonesteel2005}. Since the result of braiding operations depends only on the topological properties of the braids, not the precise details of the braiding paths, these logical control operations are intrinsically robust against noisy control operations. For certain phases with a sufficiently rich topological order, braiding operations form a universal set of quantum gates~\cite{Freedman2002a,Kitaev2003,Mochon2004,Mochon2003}. 

To robustly physically realize such a robust topological phase, it is essential to maintain equilibrium with a low temperature external reservoir that can remove entropy and any associated accumulation of excitations due to environmental noise and/or noisy control operations~\cite{Brown2007,Kitaev2006a,Dennis2002,Nayak2008,Alicki2009a,Alicki2007a}. A generic quantum simulation (e.g., with trapped cold atoms) is an open quantum system that is not intrinsically in equilibrium with an external thermal reservoir. Thus maintaining thermal equilibrium at low temperatures requires explicitly generating an effective coupling to an external reservoir. It is important to note there that while theoretical studies have shown that in the thermodynamic limit, topological order is destroyed at any finite temperature \cite{Nussinov2008a}, in a finite sized system there is a finite temperature crossover below which the topological order is preserved \cite{Castelnovo2007a}. 

Kitaev has introduced a class of exactly soluble lattice models with topologically ordered ground states~\cite{Kitaev2003,Bravyi1998,Wen2003}. The Hilbert space of these systems is defined in general by a set of qudits that sit on the links of an oriented square lattice. When this model is placed on a lattice on a torus, this is known as the toric code Hamiltonian.  We outline here the generic model of the abelian toric code as well as its non-abelian generalizations.  Each qudit state is labeled by an element of a finite group $G$, such that the local Hilbert space on each link is $\left\{ \ketil{g}, g \in G \right\}$. For each qudit, we define the operators that perform left and right multiplication
\begin{align}
L_+^h \ket{g} = \ket{hg} \quad L_-^h \ket{g} = \ket{gh^{-1}},
\end{align}
as well as the projection operators
\begin{align}
T_+^h \ket{g} = \delta_{h,g} \ket{g} \quad T_-^h \ket{g} = \delta_{h-1,g} \ket{g}.
\end{align}

The toric code Hamiltonian is a stabilizer-like Hamiltonian comprising commuting 4-body interactions:
\begin{align}
H_G = -\sum_v A_v - \sum_p B_p,
\end{align}
where the $A_v$'s are 4-body interactions defined on the vertices of the lattice and the $B_p$'s are 4-body interactions defined on the plaquettes of the lattice. The vertex terms are given by
\begin{align}
A_v = \frac{1}{\left \vert G \right \vert} \sum_{g \in G} {L_+^g}_i {L_+^g}_ j {L_-^g}_k {L_-^g}_l,
\label{eq:A_nonabelian}
\end{align}
where $\{i,j,k,l \} \in v$ and are ordered clockwise around $v$. The vertex operators can be viewed as gauge transformations, and eigenvalue $1$ eigenstate of $A_v$ are considered gauge invariant. Any state for which $A_v \ketil{\Psi_0 } \neq \ketil{ \Psi_0 }$ on a given vertex is therefore not gauge invariant, and considered to have a non-trivial electric charge at vertex $v$. 

The plaquette operators are given by:
\begin{align}
B_p = \sum_{ g_1 g_2 g_3 g_4 = \mathrm{1} } {T_-^{g_1}}_i {T_-^{g_2}}_ j {T_+^{g_3}}_k {T_+^{g_4}}_l,
\label{eq:B_nonabelian}
\end{align}
where the sum over $\{ g_1,g_2,g_3,g_4 \}$ whose product is the identity. Any state which is not a $1$ eigenvalue eigenstate of some $B_p$ is considered to have a non-trivial magnetic charge on the face of plaquette $p$.

All the $A_v$ and $B_p$ commute, and so the ground state of $H_G$ is a simultaneous eigenstate of all $A_v$ and $B_p$ operators with eigenvalue 1:
\begin{align}
A_v \ket{\Psi_0 } = \ket{ \Psi_0 }, \quad B_p \ket{ \Psi_0 }  =  \ket{ \Psi_0 }.
\end{align}
The ground state has no electric and magnetic charges, and there is a finite gap to electric and magnetic quasiparticle excitations. The stabilizer-like form of $H$ means that all excited states have purely localized electric or magnetically charged quasiparticle excitations. The spectrum of $H$ includes electric charges on the vertices, magnetic charges on the plaquettes, and bound dyonic state of electric and magnetic charges.

Magnetic charges are labeled by the conjugacy classes of $G$, where the conjugacy class $C_h$ of element $h$ is defined as
\begin{align}
C_h \equiv \left \{ ghg^{-1}; g \in G \right \}.
\end{align}
The centralizers of each conjugacy class $C_h$, which commute with all elements of $C_h$ label the electric charges. Pairs of neighboring excitations can be created by applying one-qudit operators to a link. For example, to create a pair of magnetic fluxes on neighboring plaquettes, we can define an operator $E^+_{p,p'} ([h])$ which acts on the link connecting the plaquettes $p$ and $p'$~\cite{Brennen2009b}:
\begin{align}
\ket{ m_p^{[h]}, m_{p'}^{[h]} } = E_{p,p'}^+ ([h]) \ket{ \Psi_0 }, \quad E_{p,p'}^+ ([h]) \equiv \frac{1}{\sqrt{\vert [h] \vert}} \sum_{h \in [h]} L_-^h,
\end{align}
\noindent
where $\ketil{ m_p^{[h]}, m_{p'}^{[h]} }$ is a state with magnetic fluxes of charge $[h]$ on plaquettes $p,p'$. The operator $E^+_{p,p'} ([h])$ is a one qudit operator acting on the link connecting $p$ and $p'$. An arbitrary one-qudit operator applied to the ground state will crate a superposition of electric and magnetic charges at the neighboring vertices and plaquettes.

\subsection{Abelian Toric Code}
\label{subsec:abelianTC}
The simplest choice of $G=Z_2$ results in Kitaev's canonical toric code~\cite{Kitaev2003,Bravyi1998}, which has an abelian $Z_2$ topologically ordered ground state. The $Z_2$ toric code has a 4-fold degenerate ground state on the torus. Since the braiding statistics are abelian in this case, the $Z_2$ toric code can act only as a topologically protected quantum memory and not as a universal topological computer~\cite{Dennis2002,Kitaev2003}. 

The Hamiltonian of the $Z_2$ toric code is given by
\begin{equation}
H_{\mathrm{TC}} = -\lambda_e \sum_v A_v - \lambda_m \sum_p B_p  \label{eq:H_TC}, \quad A_v \equiv \prod_{j \in v} \sigma^z_j, \quad B_p \equiv  \prod_{j\in p} \sigma^x_j,
\end{equation}
where $\{\sigma_j\}$ are a set of qubits located on the links of a square lattice on a torus, and $\sigma_j^\alpha$ is the $\alpha \in \{x,y,z\}$ Pauli operator on qubit $j$. $A_v$ and $B_p$ are the 4-body interactions around the vertices $v,$ and plaquettes $p$ of the lattice, respectively. The ground state of $H_{\mathrm{TC}}$ is a quantum liquid state in which all local correlation functions decay exponentially. Nevertheless, the ground state shows $Z_2$ topological order, manifested in the four-fold ground state degeneracy of $H_{\mathrm{TC}}$ on the torus. These degenerate ground states may be distinguished by the action of the following non-local loop operators:
\begin{equation}
  W^x_{1,2} \equiv \prod_{j \in c_{1,2}} \sigma^x_j, \quad W^z_{1,2} \equiv \prod_{j \in \tilde{c}_{1,2}} \sigma^z_j.
\end{equation}
Here $c_{1,2}$ are loops through the vertex lattice that wind around one of the two directions of the torus, and $\tilde{c}_{1,2}$ are loops passing through the faces of the plaquettes. Since $[ W^{x,z}_{1,2}, H_{\mathrm{TC}} ] = 0$, the ground states may also be written as eigenstates of $W^{x,z}_{1,2}$ with eigenvalues $\pm1$.

The excited eigenstates with $A_v \ket{\Psi_{ex}} = - \ket{\Psi_{ex}}$ have a localized ``electric" $e$-type quasiparticles on the vertex $v$ that costs energy $2 \lambda_e$.  Eigenstates with $B_p \ket{\Psi_{ex}} = - \ket{\Psi_{ex}}$ have a localized ``magnetic" $m$-type quasiparticles on the plaquette $p$ that costs energy $2 \lambda_m$. Pairs of quasiparticles can be created by applying one-body Pauli operators to the ground state:
\begin{align}
\ket{e_v e_{v'}} = \sigma^x_{v,v'} \ket{\Psi_0^{\mathrm{TC}}}, \quad \ket{m_p m_{p'}} = \sigma^z_{p,p'} \ket{\Psi_0^{\mathrm{TC}}}
\end{align}
where $\sigma_{v,v'}$ is spin on the link connecting $v$ and $v'$ and $\sigma_{p,p'}$ connects plaquettes $p$ and $p'$. The lowest energy excited states are characterized by such pairs of localized quasiparticles and are separated from the ground state by an energy gap of  $\Delta_{e,m} = 4\lambda_{e,m}$. An arbitrary one-body operator acting on an eigenstate of $H_{\mathrm{TC}}$ will act to either create or destroy pairs of quasiparticles around the link: sequential action of one-body operators therefore results in translation of a single quasiparticle across the link. Both $e$ and $m$ quasiparticles act as bosons under exchange amongst their own type: however, braiding an $e$ around an $m$ generates a phase of $-1$, and so the two types of quasiparticles are seen to have mutual abelian \emph{semionic} braiding statistics. 

On a finite-sized lattice of linear dimension $L$, the crossover temperature below which the topological order is preserved for the $Z_2$ toric code is given by $T^* \sim \Delta/ \ln{L}$ \cite{Castelnovo2007a}. The anyonic braiding statistics allow this abelian toric code to therefore act as the basis for a robust quantum memory as long as it is kept in equilibrium at a low temperature, $T \ll T^*$.  However, as noted above, the abelian topologically protected braiding operations are not universal for computation in the $Z_2$ model.

\subsection{Nonabelian Toric Code}
\label{sub:nonabelian_toric_code}

Mochon \cite{Mochon2003,Mochon2004} has shown that for certain non-abelian finite groups $G$, the braiding and fusion of electric and magnetic fluxes can lead to formation of a universal logical gate set. Thus, if the corresponding toric code Hamiltonians $H_G$ can be robustly generated and localized quasiparticles controllably created and manipulated, the non-Abelian toric code may act as the basis for a topologically fault tolerant quantum computer. The smallest non-abelian group is the group of permutations of three objects, $S_3$ and this would already allow for a topologically protected universal gate set~\cite{Mochon2003,Mochon2004}. Since $S_3$ has 6 elements, simulating $H_{S_3}$ therefore requires using qudits with $d=6$ at each link of the square lattice. In Section~\ref{sec:simulating_many_body_quantum_operators} we indicate how the corresponding vertex and plaquette operators of Eqs.~(\ref{eq:A_nonabelian})-(\ref{eq:B_nonabelian}) may be constructed from a universal set of one and two-qudit gates. 


\subsection{Finite Temperature Behavior}

As noted above, while topological order is unstable at any finite temperature in the thermodynamic limit, in a finite sized system there is a finite crossover temperature below which the topological order may be preserved~\cite{Castelnovo2007a,Nussinov2008a}. To keep the system in such effective low temperature state with respect to $H_G$, or to cool all the way to the ground state, we must therefore couple each vertex and plaquette to a set of ancillary reservoir qudits undergoing dissipation~\cite{Alicki2009a,Alicki2007a}.  In the next section we describe how this thermalization may be carried out in an efficient manner, focusing on the qubit case, i.e., on the abelian $Z_2$ toric code. 


\section{Thermalization of stabilizer Hamiltonians} 
\label{sec:thermalization_of_toric_code}

Since noisy control operations and interactions with the environment will introduce entropy to the system, a robust quantum simulation requires a dissipative process to remove entropy and effectively cool the system~\cite{Brown2007,Lloyda}. As an alternative to algorithmic error correction, we present an approach here to maintain the system in equilibrium with an effective thermal reservoir. If the effective temperature of the reservoir approaches zero, this thermalization will act purely as passive error correction; however, our approach also provides for the ability to tune the temperature of the reservoir and equilibrate the system at finite temperature. In addition to providing robustness, this thermal equilibration also therefore would allow for a study of thermal properties of the quantum model at hand.
One procedure for extracting finite temperature properties of a many-body system is to simulate its thermalization by coupling it to external dissipative modes. The steady-state of the following Lindblad master equation is guaranteed to produce the thermal equilibrium density matrix of the system:
\begin{equation}
\frac{\dt \rho}{\dt{t}} = -\frac{i}{\hbar}[H_0,\rho] + \sum_k \gamma_k \mathcal{D}[H^-_k] \rho + \sum_k e^{-\beta \hbar \omega_k}\gamma_k \mathcal{D}[H^+_k]\rho 
\label{eq:thermal_me}
\end{equation}
where $\mathcal{D}[A]\rho \equiv 2 A \rho A\dg - A\dg A\rho - \rho A\dg $, and $H^+_k$ creates an excitation of energy $\hbar\omega_k$. The unique steady state of this master equation is the thermal density matrix corresponding to the inverse temperature $\beta$ and Hamiltonian $H_0$ \cite{Breuer2002}. Despite this simple description, the mathematical model above can be difficult to simulate in practice because the excitation creation and annihilation operators $H^+_k, H^-_k$ are usually a superposition of non-local many-body operators in an interacting system. For the abelian toric code these are $4$-body operators, and therefore each generator in the Lindblad terms is a $4$-body term. 

These $4$-body operators can be implemented using the stroboscopic technique described in Ref. \cite{Herdman2010c}.  However such an implementation is resource intensive.  Here we describe a simpler implementation of thermalization in stabilizer Hamiltonians by exploiting the local properties of excitations. Our analysis in this section focuses on stabilizer Hamiltonians of qubits for simplicity, however the arguments presented here may be extended to qudit Hamiltonians, as required for realization of, e.g., a non-abelian toric code.  The key insight for constructing a simple thermalization scheme is to note that a local Pauli operation in the interaction picture defined by the stabilizer Hamiltonian (an eigenoperator decomposition of the local Pauli) decomposes into a small number of Fourier components:
\begin{equation}
	e^{\frac{i}{\hbar}t H^\textrm{stab}} \sigma^{\alpha}_j e^{-\frac{i}{\hbar}t H^\textrm{stab}} 
		= \sum_{k=1}^M e^{-\frac{i}{\hbar}2\epsilon_{\alpha,j,k} t}a_{\alpha,j,k} + e^{\frac{i}{\hbar}2\epsilon_{\alpha,j,k} t}a\dg_{\alpha,j,k} 
\label{eq:fourier_general}
\end{equation}
where $\alpha \in \{x,y,z\}$, $j$ indexes a qubit in the many-body system, and $a_{\alpha,j,k}, a_{\alpha,j,k}\dg$ are generalized creation and annihilation operators for excitations of energy $\epsilon_{\alpha,j,k}$, associated with the Pauli operator $\sigma^\alpha_j$. While for a general Hamiltonian this eigenoperator decomposition can have a number of terms that is extensive in system size, the number of terms above, $M$, will necessarily be small because the terms in $H_\textrm{stab}$ commute with themselves and only a small number of them do not commute with the local Pauli operator $\sigma^\alpha_j$. 

In our construction we will interact each local qubit with up to $M$ ancillary spins that are driven to a thermal state. The specific form of the interaction Hamiltonian is:
\begin{equation}
	H_\textrm{int} 
		= \sum_{j=1}^N \sum_{\alpha} \sum_{k=1}^M \sigma^\alpha_j \otimes \Sigma^x_{\alpha,j,k} 
\end{equation}
where $\Sigma^x_{\alpha,j,k}$ is the Pauli $x$ operator on ancilla spin $k$ for the $\alpha$th Pauli operator on qubit $j$ in the stabilizer Hamiltonian. Note that this interaction Hamiltonian is two-body and each lattice site interacts with at most $M$ ancillas.. The Zeeman frequency of the $k$th ancilla spin is chosen to be $\omega_{\alpha,j,k} = \epsilon_{\alpha,j,k}/\hbar$. 

The effective dynamics of the system and ancillary spins, with combined density matrix $\Omega \in \mathcal{H}^{\textrm{lattice}}\otimes \left(\mathbb{C}^2\right)^{\otimes MN}$, is given by:
\begin{align}
	\frac{\dt{\Omega}}{\dt{t}} 
		&= -\frac{i}{\hbar}[H^\textrm{stab} 
			- \sum_{\alpha, k, j} \hbar\omega_{\alpha,j,k} \Sigma^{z}_{\alpha,j,k} 
		         + H_\textrm{int}, \Omega] \nn \\
		& + \sum_{\alpha,j,k} \gamma_{\alpha,j,k}^-\mathcal{D}[\Sigma^{-}_{\alpha,j,k}]\Omega 
			+ \sum_{\alpha,j,k} \gamma_{\alpha,j,k}^+ \mathcal{D}[\Sigma^{+}_{\alpha,j,k}]\Omega
\end{align}
The dissipation rates for all the ancillary spins are chosen such that 
\begin{equation}
	\frac{\gamma^-_{\alpha,j,k}}{\gamma^+_{\alpha,j,k}} 
		= e^{-\beta \hbar \omega_{\alpha,j,k}}
\end{equation}
so that in the absence of the coupling to stabilizer Hamiltonian qubits, the unique steady state of dissipative dynamics of a single ancillary spin is the thermal density matrix at inverse temperate $\beta$: 
\begin{equation}
\left(\begin{array}{cc}p^0_{\alpha,j,k} & 0 \\0 & p^1_{\alpha,j,k}\end{array}\right)
\end{equation}
with $p^1_{\alpha,j,k} = 1-p^0_{\alpha,j,k} = e^{-\beta \hbar \omega_{\alpha,j,k}}p^0_{\alpha,j,k}$.
Such a thermally driven ancilla system can be implemented as a driven three-level lambda-configuration atom with a strong spontaneous emission channel, resulting in the two stable levels having thermal population distribution; in such a case they are pseudospins. The inverse temperature of the ancilla pseudospins is chosen to be $\beta = 1/k_BT$ (where $T$ is the desired temperature for the simulation).

Transforming into the interaction picture with respect to $H^\textrm{stab} - \sum_{\alpha,j,k}  \hbar\omega_{\alpha,j,k}  \Sigma^{z}_{\alpha,j,k}$, and using \erf{eq:fourier_general} yields an interaction picture interaction Hamiltonian of the form:
\begin{align}
\tilde{H}_\textrm{int} 
		&= \sum_{j=1}^N \sum_{\alpha} \sum_{k=1}^M \left( \sum_{l=1}^M e^{-\frac{i}{\hbar}2\epsilon_{\alpha,j,l} t}a_{\alpha,j,l} + e^{\frac{i}{\hbar}2\epsilon_{\alpha,j,l} t}a\dg_{\alpha,j,l} \right) \otimes \left( \Sigma^-_{\alpha,j,k}e^{-i2\omega_{\alpha,j,k}t} +  \Sigma^+_{\alpha,j,k}e^{i2\omega_{\alpha,j,k}t} \right)
\end{align}
Remembering that  $\hbar\omega_{\alpha,j,k} = \epsilon_{\alpha,j,k}$ and using the rotating wave approximation to drop all oscillating terms reduces the interaction Hamiltonian to
\begin{equation}
\tilde{H}^{\textrm{RWA}}_\textrm{int} =\sum_{j=1}^N \sum_{\alpha} \sum_{k=1}^M a_{\alpha,j,k}\otimes\Sigma^+_{\alpha,j,k} + a\dg_{\alpha,j,k} \otimes \Sigma^-_{\alpha,j,k}
\end{equation}
Therefore the effective evolution in the interaction picture and under the rotating wave approximation is given by the master equation:
\begin{align}
	\frac{\dt{\Omega}_\textrm{int}}{\dt{t}} 
		&= -\frac{i}{\hbar}[\sum_{j=1}^N \sum_{\alpha} \sum_{k=1}^M a_{\alpha,j,k}\otimes\Sigma^+_{\alpha,j,k} + a\dg_{\alpha,j,k} \otimes \Sigma^-_{\alpha,j,k}, \Omega] \nn \\
		& + \sum_{\alpha,j,k} \gamma^-_{\alpha,j,k}\mathcal{D}[\Sigma^{-}_{\alpha,j,k}]\Omega 
			+ \sum_{\alpha,j,k} \gamma^+_{\alpha,j,k} \mathcal{D}[\Sigma^{+}_{\alpha,j,k}]\Omega
\label{eq:me_rwa1}
\end{align}
This evolution describes energy exchange of the stabilizer Hamiltonian system with a set of thermalized ancilla spins, and will result in the thermalization of the system. Essentially the engineered dissipation channels provided by the ancilla pseudospins mimic a bath satisfying detailed balance at the resonance frequencies of the system. Although we can demonstrate the thermalization of the system in this general setting (see Appendix), for concreteness we will now illustrate it explicitly for the abelian toric code.

\begin{figure}
\begin{align*}
	E\dg_{j,\mathcal{E}} \ket{\myvcenter{\includegraphics[width=10ex]{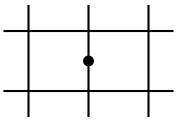}}} 
		&= \ket{\myvcenter{\includegraphics[width=10ex]{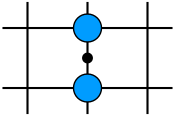}}} \\
	T_{j,\mathcal{E}} \ket{\myvcenter{\includegraphics[width=10ex]{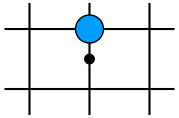}}} 	
		&= \ket{\myvcenter{\includegraphics[width=10ex]{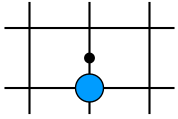}}}\\
	E\dg_{j,\mathcal{M}} \ket{\myvcenter{\includegraphics[width=10ex]{empty.png}}} 
		&= \ket{\myvcenter{\includegraphics[width=10ex]{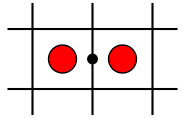}}} \\
	T_{j,\mathcal{M}} \ket{\myvcenter{\includegraphics[width=10ex]{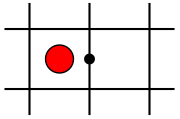}}} 	
		&= \ket{\myvcenter{\includegraphics[width=10ex]{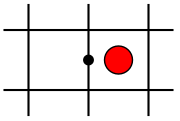}}}
\end{align*}
\caption{Graphical representation of the action of excitation creation and translation operators for the toric code. The black dot indicates site (qubit) $j$, the blue (red) circles indicate electric (magnetic) excitations. \label{fig:operator_action}}
\end{figure}

For the toric code, the Fourier decomposition of the local Pauli terms of interest is:
\begin{equation}
	e^{\frac{i}{\hbar}t H_\textrm{TC}} \sigma^{x/z}_j e^{-\frac{i}{\hbar}t H_\textrm{TC}} 
		= e^{-2\frac{i}{\hbar}\Delta t}E_{j,\mathcal{E}/\mathcal{M}} 
		+ e^{2\frac{i}{\hbar}\Delta t}E_{j,\mathcal{E}/\mathcal{M}}\dg 
		+ T_{j,\mathcal{E}/\mathcal{M}}
\label{eq:localexc_toric}
\end{equation}
$\nu=\mathcal{E}$ and $\nu=\mathcal{M}$ denote electric and magnetic excitations, and we assume the electric and magnetic excitations have the same energy: $\lambda_e = \lambda_m = \Delta$. The operator $E\dg_{j,\nu}$ creates a pair of excitations of type $\nu$ about site $j$ and $T_{j,\nu} = T\dg_{j,\nu}$ translates a type-$\nu$ excitation about site $j$. See Ref. \cite{Herdman2010c} for the formal definition of these operators, and see Fig. \ref{fig:operator_action} for graphical representation of the action of these operators.  In terms of these operators, the toric code Hamiltonian may be written as,
\begin{equation}
\label{eq:HTCfourier}
	H_{\rm TC} = \frac{\Delta}{8} \sum_{j,\nu} \left(2E_{j,\nu}^\dagger E_{j,\nu} + T_{j,\nu}^2\right) 	
\end{equation}

This decomposition of toric code excitations was used by Alicki, Fannes, and Horodecki \cite{Alicki2009a,Alicki2007a} to show that thermalization of the toric code is guaranteed under a local interaction Hamiltonian coupling to a thermal bath of the form:
\begin{equation}
	H_\textrm{int,harmonic} 
		= \sum_{j=1}^N \sigma^x_j \otimes f_j + \sum_{1}^N \sigma^z_j \otimes g_j
\end{equation}
where $f_j, g_j$ are Hermitian bath operators that satisfy detailed balance. For example, in the case of a bath of free harmonic modes, $f_j, g_j$ could be the sum over displacement operators for the modes. We will instead, show that by the above construction one can also simulate thermalization by utilizing ancillary pseudospins as the engineered dissipative environment. The interaction Hamiltonian we require for the toric code example is
\begin{equation}
	H_\textrm{int} 
		= \sum_{j=1}^N \sigma^x_j \otimes  \Sigma^x_{\Delta, j, \mathcal{E}} + \sigma^z_j \otimes  \Sigma^x_{\Delta, j, \mathcal{M}}  + \sigma^x_j \otimes  \Sigma^x_{0, j, \mathcal{E}} + \sigma^z_j \otimes  \Sigma^x_{0, j, \mathcal{M}} 
\end{equation}
Here, the $\Sigma^x_{\Delta/0, j, \mathcal{E/M}}$ are Pauli-$x$ operators on ancillary pseudospins which are implemented as driven three-level lambda-configuration atoms with strong spontaneous emission. Each toric code lattice spin has four types of ancillary pseudospin coupled to it ($\Sigma_{\Delta,\mathcal{E}}, \Sigma_{0, \mathcal{E}}, \Sigma_{\Delta, \mathcal{M}}$ and $\Sigma_{0, \mathcal{M}}$). The characteristic frequency of all $\Delta$-type ancillary pseudospins is $\omega_\Delta = \Delta/\hbar$, and all $0$-type ancillary pseudospins is $\omega_0 = 0$ (these correspond to the two eigenfrequencies in the decomposition \erf{eq:localexc_toric}). Transforming into the interaction picture with respect to $H_\textrm{TC} - \sum_{j,\nu} \hbar\omega_\Delta \Sigma^{z}_{\Delta, j, \nu} - \sum_{j} \hbar\omega_0 \Sigma^{z}_{0, j, \nu}$, with $\nu \in \{\mathcal{E}, \mathcal{M} \}$, and employing the rotating wave approximation produces:
\begin{equation}
\tilde{H}^{\textrm{RWA}}_\textrm{int} = \sum_{j,\nu} E_{j,\nu}\otimes \Sigma^+_{\Delta, j, \nu} + E_{j,\nu}\dg \otimes \Sigma^-_{\Delta, j, \nu} +  T_{j,\nu} \otimes \Sigma^x_{0, j, \nu}
\end{equation}

Therefore the effective evolution in the interaction picture and under the rotating wave approximation is given by the master equation:
\begin{align}
	\frac{\dt{\Omega}}{\dt{t}} 
		&= -\frac{i}{\hbar}[\sum_{j,\nu} E_{j,\nu}\otimes \Sigma^+_{\Delta, j, \nu} + E_{j,\nu}\dg \otimes \Sigma^-_{\Delta, j, \nu} +  T_{j,\nu} \otimes \Sigma^x_{0, j, \nu}, \Omega] \nn \\
		& + \sum_{j,\nu} \gamma^-_\Delta \mathcal{D}[\Sigma^{-}_{\Delta, j, \nu}]\Omega 
			+ \sum_{j} \gamma^+_\Delta \mathcal{D}[\Sigma^{+}_{\Delta, j, \nu}]\Omega 
			+ \sum_{j} \gamma^-_0 \mathcal{D}[\Sigma^{-}_{0, j, \nu}]\Omega 
			+ \sum_{j} \gamma^+_0 \mathcal{D}[\Sigma^{+}_{0, j, \nu}]\Omega 
			\equiv \mathcal{L}_\textrm{eff}\Omega
\label{eq:me_rwa2}
\end{align}
If we choose $\gamma^+_\Delta = e^{-\beta_\mathsf{T}\Delta}\gamma^-_\Delta$ and $\gamma^+_0 = \gamma^-_0$, which results in the $\Delta$-type ancillary pseudospins thermalizing to a finite inverse temperature $\beta$ and the $0$-type ancillary pseudospins thermalizing to infinite temperature ($\beta=0$), it is shown in the Appendix that the unique steady state of this master equation is the thermal state of the combined system at inverse temperature $\beta$. That is, 
\begin{equation}
\mathcal{L}_\textrm{eff} ~\frac{e^{-\beta H_\textrm{TC}}\otimes e^{-\beta \left( - \sum_{j,\nu} \hbar\omega_\Delta \Sigma^{z}_{\Delta, j, \nu} - \sum_{j,\nu} \hbar\omega_0 \Sigma^{z}_{0, j, \nu}\right)}}{\mathcal{Z}} = 0
\label{eq:toric_ss}
\end{equation}
and this is the only state that satisfies this property.

The above analysis for simulating thermalization focused explicitly on qubit stabilizer Hamiltonians. However, the same construction follows for qudit stabilizer Hamiltonians because the critical property of local excitations and a decomposition analogous to \erf{eq:fourier_general} also holds for these. In the qudit case however, there are more than three Pauli generators in the group of local generators and thus the Fourier decomposition will be more involved. As a consequence the number of ancilla spins necessary to simulate the thermal bath will also increase.

\section{Physical simulation of stabilizer system} 
\label{sec:simulating_many_body_quantum_operators}

We shall now consider the physical context in which the aforementioned thermal stabilizer system is to be simulated.  Though a number of proposals exist for quantum simulation, for specificity we shall restrict our discussion to arrays of trapped neutral atoms.  In particular, we consider a set of $\sim 250$ individual $^{133}$Cs atoms trapped at the sites of an addressable, simple cubic optical lattice~\cite{Nelson2007a}.   The orbital degrees of freedom are slow, and can be considered as effectively frozen on the time scales relevant to our analysis.  We therefore need consider only the internal atomic degrees of freedom, from which we select two hyperfine levels (e.g., $\ket{F,m_F} = \ket{4,4}, \ket{3,3} $) to define a 2-level pseudospin system.  The Hamiltonian, \erf{eq:H_TC}, will be implemented in an interaction picture with respect to the atomic energy levels. Additionally, we choose auxiliary internal levels to serve as intermediate states to facilitate optical frequency Raman transitions for single qubit operations, as well as a highly excited, $n\simeq 80$, Rydberg level necessary for two qubit interactions~\cite{Jaksch2000, Cozzini2006}.  Since the pseudospins are localized at the sites of a cubic lattice, one can choose to either realize the dynamics on a single plane using a surface code ~\cite{Bravyi1998, Freedman2001} or in a three-dimensional cubic array with toroidal boundary conditions realized by SWAP operations.   In addition to this set of system qubits on which the stabilizer Hamiltonian is simulated, we must include a set of ancillary atoms to serve as a thermal reservoir.  This ancillary set must be strongly dissipative and will be optically pumped to produce the desired thermal state.

\begin{figure}[ht]
\begin{center}
\label{fig:circuit}
\includegraphics[width=.8\columnwidth]{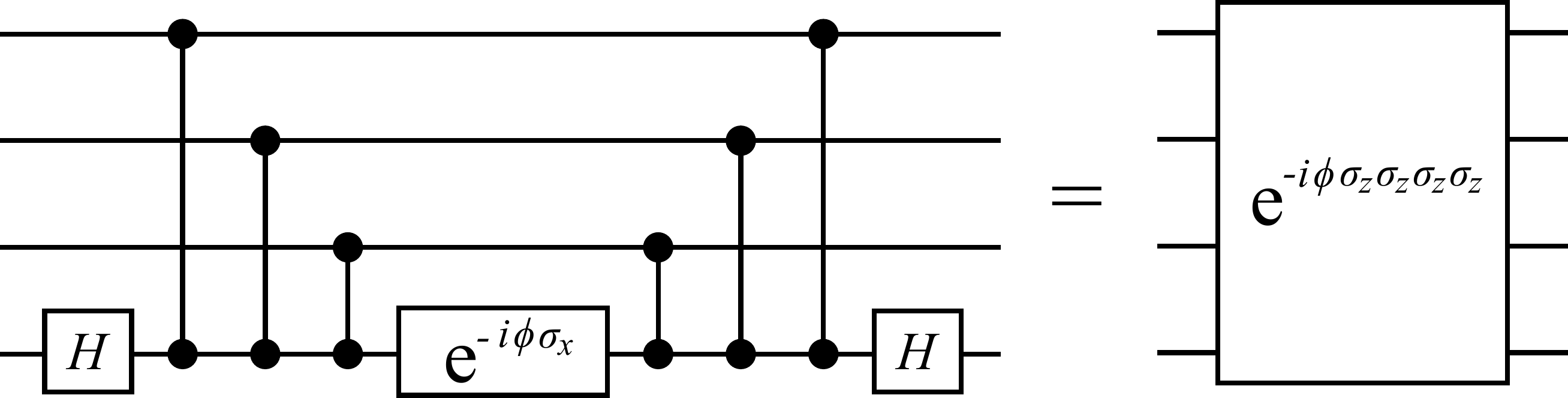}
\caption{Ancilla-free quantum circuit nonperturbatively implementing action of many body Hamiltonian.  The circuit on the left utilizes gates available to the neutral atom array described in the text.  Vertical lines indicate Rydberg CPHASE gates between qubits indicated by black dots and $H$ indicates a Hadamard rotation.  Action of Hamiltonian proportional to $\sigma_x\sigma_x\sigma_x\sigma_x$ may be implemented by conjugation of the above circuit with Hadamard operations.}
\end{center}	
\end{figure}

The analysis of the previous section demonstrates that the thermal properties of stabilizer Hamiltonians may be studied by coupling to a dissipative bath of two-level systems.  However, it is not possibly to directly implement the Liouvillian on the neutral atom system, so we instead take a stroboscopic approach.  This approach applies a sequence of local operations to generate an evolution which closely approximates the evolution generated by the dynamical equations.  We shall begin by considering the generic Lindblad master equation,
\begin{equation}
\label{eq:liouvmaster}
	\dot\rho = \mathcal{L}\rho(t) = \sum_j \mathcal{A}[H_j]\rho(t) + \sum_k\gamma_k\mathcal{D}[K_k] \rho(t).
\end{equation}
Here we have used 
	$A[H_j]\rho(t) = -i [H_j, \rho(t)]/\hbar$  
to represent the adjoint action of $H_j$.  Evolution under this equation generates the time evolution operator,
\begin{equation}
\label{eq:Loft}
	L(t) = \exp(\mathcal{L} t)
\end{equation}
The master equation, \erf{eq:liouvmaster}, is impossible to implement continuously in the neutral atom array, so we approximate the evolution by a sequence of local operators using a Trotter expansion, 
\begin{equation}
\label{eq:trotter}
	L(t) = \exp\left(\order{1/N^2}\right)\left(\prod_j\exp\left(\mathcal{A}[H_j] \Delta t\right)\prod_k\exp\left(\gamma_k\mathcal{D}[K_k] \Delta t\right)\right)^N,
\end{equation}
where $\Delta t = t/N$.By choosing a sufficiently large $N$, the error term can be made arbitrarily small.  Each term in the master equation may then be simulated independently over the short time $\Delta t$.  For a generic stabilizer Hamiltonians, however, many of these terms will be multibody.  Couplings derived from first principles physical interactions, on the other hand, are intrinsically two-body.  Exceptions to this are rare, and usually derive from an implicit averaging over time or intermediate degrees of freedom \cite{Weimer2010}. In Ref.~\cite{Herdman2010c}, we presented a perturbative method based on the Magnus expansion for simulating many-body interactions between qubits by application of a sequence of two-body quantum gates.  Here we present a non-perturbative method for the simulation of arbitrary $n$-body qubit gates based on a construction in Ref.~\cite{mikeandike, Barenco1995}.  This method displays significantly lower overhead in terms of total gate count as compared to the Magnus expansion approach. While the method is capable of simulating a generic product of Pauli matrices, we shall explicitly consider here only the terms necessary for simulation of the abelian toric code Hamiltonian \erf{eq:H_TC}.  The Trotter expansion leaves us to simulate unitary evolution operators of the form $U_Z(t) = \exp(-i \phi \sigma_z\sigma_z\sigma_z\sigma_z)$, with $\phi = \lambda \Delta t$. This evolution may be simulated by the sequence of CPHASE and one-qubit gates shown in Fig.~\ref{fig:circuit}.  This circuit may be readily extended by single qubit operations to achieve any product of four Pauli matrices.  For example, pre- and post- multiplying by Hadamard transformations produces $U_X = \exp(-i\phi \sigma_x\sigma_x\sigma_x\sigma_x) = H^{\otimes 4} U_Z H^{\otimes 4}$.  

Simulation of the dissipative terms in \erf{eq:liouvmaster} may be effected through the use of atomic states with strong spontaneous emission.  This dissipation forms the entropy reducing part of the map.  As shown in the Appendix, the stationary state of the evolution is thermal regardless of the magnitude of the dissipation couplings, $\gamma_k$, so long as they are positive.  We may therefore choose them to be arbitrarily large, effectively replacing the operators $\prod_k\exp(\mathcal{D}[K_k] \delta_t)$ by a reset to the thermal state.  This could be accomplished in a number of ways.  For example, correctly tuned coherent driving to a state with strong dissipation can optically pump the atom directly to the thermal state.  Alternatively, one may measure the ancilla bit in the computational basis and apply a $\pi$-pulse conditioned on the result of the measurement and a classical, Boltzmann-weighted random bit.  

For the non-abelian toric code we need to generate the generalized 4-body interactions in \erf{eq:A_nonabelian}.  This requires an implementation of the $S^{3}$ multiplication table on $\mathbb{C}^{6}$, the single qudit Hilbert space. Given an implementation of one- and two-qudit gates, we can generate the required 4-body terms by methods similar to those above for the qubit case.  Brennen \emph{et al} \cite{Brennen2005} have demonstrated the existence of universal sets of one and two qudit gates. It remains to construct explicit instances of gate sequences. Arbitrary one-qudit gates may be implemented through entirely local unitary actions on a six level qudit pseudospin system. In the case of trapped neutral atoms, this is possible with methods similar to the single qubit unitaries.  For a pair of levels within one six level system and a corresponding pair in one of its neighbors, we may construct a Rydberg blockade, analogous to the qubit case.  This allows the explicit construction of two qudit gates. Details of this will be discussed in a forthcoming paper.


\section{Summary}
\label{sec:summary}
Motivated by the need to incorporate dissipative processes to remove entropy and provide cooling during quantum simulations, we have developed a scheme for efficient finite temperature quantum simulation of general stabilizer Hamiltonians.  These Hamiltonians are typically characterized by non-local or many-body interactions that are hard to realize.  The well known toric code Hamiltonian of Kitaev, which allows both abelian representations with qubits and non-abelian representations with qudits, is taken here as a canonical example of stabilizer Hamiltonian in order to demonstrate the approach. Our method relies on coupling of each physical qubit or qudit involved in the Hamiltonian simulation to a small number of ancillary pseudospins that are dissipatively driven to reach a specific temperature.  By using a Fourier decomposition of local Pauli operations on the physical qubits, we show that we can achieve thermalization by employing only two-body couplings between the physical qubits with a small number of ancillary pseudospins.  This is a significant improvement over our earlier work ~\cite{Herdman2010c}, which required that the dissipatively driven ancillas be coupled to the physical qubits by similar many-body interactions as those contained in the stabilizer Hamiltonians.  We illustrated the thermalization approach explicitly for the abelian toric code of Kitaev, where two-body interactions with two dissipatively driven ancilla pseudospins are all that is required to achieve thermalization of a finite set of qubits evolving under the toric code Hamiltonian. This considerably simplifies the physical implementation with neutral atoms trapped in optical lattices, for which we also presented an improved approach to quantum simulation of the toric code Hamiltonian.  The approach can readily be extended to thermalization of quantum simulations with general stabilizer Hamiltonians, in particular to the simulation of the non-abelian toric code Hamiltonian, for which the smallest qudit dimension is six.  Detailed analysis of such non-abelian simulations with trapped neutral atoms will be presented elsewhere.

\section{Acknowledgments} 
\label{sec:acknowledgments}
This material is based upon work supported by DARPA under Award No. 3854-UCB-AFOSR-0041.  Sandia National Laboratories is a multi-program laboratory managed and operated by Sandia Corporation, a wholly owned subsidiary of Lockheed Martin Corporation, for the U.S. Department of Energy's National Nuclear Security Administration under contract DE-AC04-94AL85000.


\subsection*{Appendix A: Thermal state fixed by evolution}
\label{ap:fixed}
Here we will show that the thermal state of the abelian toric code is the unique fixed point of the engineered dissipative evolution detailed in section \ref{sec:thermalization_of_toric_code}. Consider in particular, the evolution prescribed by \erf{eq:me_rwa2}. The dissipative dynamics for the ancillary pseudospins is particularly simple since it is incoherent excitation at rate $\gamma^+$ and damping at rate $\gamma^-$ of each pseudospin independently. The \textit{only} steady-state of this evolution is the mixed state of each pseudospin 
\begin{equation}
\left(\begin{array}{cc}p^0_{\Delta/0} & 0 \\0 & 1-p^0_{\Delta/0}\end{array}\right)
\end{equation} 
with $p^0_{\Delta/0} = \gamma^-_{\Delta/0} / (\gamma^-_{\Delta/0} + \gamma^+_{\Delta/0})$. With the choice of rates given in the main text, this is a thermal state at inverse temperature $\beta$ for $\Delta$-type pseudospins and a completely mixed state for the $0$-type pseudospins -- i.e. $p^1_{\Delta/0}/p^0_{\Delta/0} = e^{-\beta \hbar \omega_{{\Delta/0}}}$, where $p^1_{\Delta/0} \equiv 1- p^0_{\Delta/0}$.

Hence the steady state of master equation evolution must have the form:
\begin{equation}
	\Omega_\textrm{ss} 
		= \rho 
		\otimes 
		\underbrace{\rho_{\Delta}^{\otimes 2N}}_{\Sigma_{\Delta} 
		\textrm{pseudospins}} 
		\otimes  
		\underbrace{\rho_{0}^{\otimes 2N}}_{\Sigma_{0} \textrm{pseudospins}}
\end{equation}
with additional conditions on $\rho$, the state of the toric code lattice spins. Here, 
\begin{equation}
\rho_{\Delta} \equiv \left(\begin{array}{cc} \frac{1}{1+e^{-\beta \Delta}} & 0 \\0 & \frac{1}{1+e^{\beta \Delta}}\end{array}\right) ~~~~~ \textrm{and} ~~~~~ \rho_{0} \equiv \left(\begin{array}{cc} \frac{1}{2} & 0 \\0 & \frac{1}{2}\end{array}\right)
\end{equation}
Substituting this form into the master equation results in:
\begin{equation}
\frac{\dt{\Omega_{\textrm{ss}}}}{\dt{t}} = -\frac{i}{\hbar}\sum_{j,\nu}\left[ E_{j,\nu}\otimes \Sigma^+_{\Delta, j, \nu} + E_{j,\nu}\dg \otimes \Sigma^-_{\Delta, j, \nu} +  T_{j,\nu} \otimes \Sigma^x_{0, j, \nu}, \rho\otimes \rho^{\otimes 2N}_{\Delta} \otimes \rho_{0}^{\otimes 2N}\right]
\end{equation}
since the dissipative terms all evaluate to zero. Now, expanding out the steady-state form for the ancilla pseudospins, and requiring that this time derivative be zero results in the following conditions $\forall ~j, \nu$:
\begin{align}
	\label{eq:lowering}
	p^0_{\Delta} E_{j,\nu}\rho - p^1_{\Delta} \rho E_{j,\nu} &= 0\\
	\label{eq:raising}
	p^1_{\Delta} E\dg_{j,\nu}\rho - p^0_{\Delta} \rho E\dg_{j,\nu} &= 0\\
	\label{eq:translation}
	T_{j,\nu}\rho - \rho T_{j,\nu} &= 0
\end{align}

At this point note that the translation operators $T_{j,\nu}$ generate a group whose action commutes with the Hamiltonian, $H_\textrm{TC}$, and is ergodic in each energy eigenspace, i.e., any two degenerate energy eigenstates are connected by a product of $T_{j,\nu}$ (with the exception of the groundspace, which we shall address later).  Because the translation operators commute with the Hamiltonian and are ergodic on each energy eigenspace, they may be decomposed into irreducible representations as
\begin{equation}
\label{eq:irrepT}
	T_{j,\nu} = \sum_n P_n T_{j,\nu} P_n
\end{equation}
where $P_n$ is the projector onto the $n^{\rm th}$ energy eigenspace. Furthermore,
\begin{equation}
\label{eq:rhodecompose}
	\rho = \sum_{n,m} P_n \rho P_m
\end{equation}
The commutation relationship $[T_{i,\nu}, \rho] = 0$ implies then that,
\begin{align}
	P_m [T_{i,\nu}, \rho] P_n 
		\notag
		&= P_m (T_{i,\nu} \rho - \rho T_{i,\nu} )P_n\\
		\label{eq:schurs}
		&= \left(P_m T_{i,\nu} P_m \right)\left(P_m \rho P_n\right) - \left(P_m \rho P_n \right)\left(P_n T_{i,\nu} P_n\right)
\end{align}
Choosing $m=n$, we see that $P_m \rho P_m$ commutes with all elements of the $m^{\rm th}$ irreducible representation of $T$.  By Schur's first lemma~\cite{Fulton1991}, this implies that $P_m \rho P_m$ must be proportional to the identity.  If we choose $m\ne n$, \erf{eq:schurs} and Schur's second lemma~\cite{Fulton1991} imply that $P_m \rho P_n = 0$.   

For the groundspace, we examine the commutation relations with the string operators.  The string operators can be represented as exciton pair creations, translations and annihilations.  But the ``commutation" relations \erf{eq:lowering}-\erf{eq:translation} imply that $\rho$ commutes with any product of $T$, $E$, and $E^\dagger$ for which there are equal numbers of $E$'s and $E^\dagger$'s.  Because $P_0 \rho P_0$ commutes with all of the string operators, Schur's lemma again implies that it too must be proportional to the identity.  

That the populations must satisfy the Boltzman distribution is insured by \erf{eq:lowering}, and so 
\begin{equation}
\label{eq:rhothermal}
	\mathcal{L}\rho=0 \implies \rho = \frac{\exp(-\beta H_\textrm{TC})}{\tr \exp(-\beta H_\textrm{TC})} \otimes \rho^{\otimes 2N}_{\Delta} \otimes \rho_{0}^{\otimes 2N}
\end{equation}

Given that the thermal state is the unique steady state of this Lindblad evolution, it can also be shown that it is an attractor, meaning that all states converge to it asymptotically \cite{Schirmer2010}.  
In fact, the above is an explicit demonstration of a very general statement about semigroups to be found in the work of Arveson \cite{Arveson1997} -- roughly, if a semigroup dynamics (e.g. generated by a Lindblad master equation) has an invariant state, and is ergodic, then it is the unique invariant state and is furthermore an attractor. The ergodicity of the dynamics is the key element, and for a stabilizer Hamiltonian it can be shown that if the Lindblad generators are excitation creation, annihilation and translation operators the system is ergodic. Such an argument can be used to prove that in the general qudit stabilizer Hamiltonian case a construction analogous to the one in Section \ref{sec:thermalization_of_toric_code} will fix the thermal state, and only the thermal state, of the system. This reflects a general pattern in the thermalization of stabilizer codes which we will discuss in a forthcoming paper.

\bibliography{stroboJPB}

\end{document}